\documentstyle[aps,psfig,floats]{revtex}

\begin{document}

\author{D.V.Dmitriev $^1$ , V.Ya.Krivnov $^2$ and A.A.Ovchinnikov $^2$}
\address{Max-Planck-Institut fur Physik Komplexer Systeme, 
Nothnitzer Str. 38, 01187 Dresden, Germany \\
and Joint Institute of Chemical Physics of RAS,
Kosygin str.4, 117977, Moscow, Russia.}
\title{Exactly solvable spin ladder model with degenerate ferromagnetic
and singlet states}
\maketitle

\begin{abstract}
We study the spin ladder model with interactions between spins on
neighboring rungs. The model Hamiltonian with the exact singlet ground state
degenerated with ferromagnetic state is obtained. The singlet ground state
wave function has a special recurrent form and depends on two parameters.
Spin correlations in the singlet ground state show double-spiral structure
with period of spirals equals to the system size.
For special values of parameters they have exponential decay. The spectrum
of the model is gapless and there are asymptotically degenerated 
excited states for special values of parameters in the thermodynamic limit.
\end{abstract}


\section{Introduction}

There has been growing interest lately in quantum spin systems with
frustrated interactions \cite{1}. Of special importance are models for
which it is possible to construct an exact ground state. The first example
of such a model is the well-known Majumdar-Ghosh model \cite{2} which is
the $s=1/2$ chain with antiferromagnetic interactions $J_{1}$ and $J_{2}$ of
nearest neighbor and next-nearest neighbor spins, where $J_{2}=J_{1}/2$.
Afterwards, a large class of 1D models with exact ground state has been
found and studied \cite{3}-\cite{12}.

A considerable progress has been achieved in a construction of such models
by using so-called matrix-product (MP) form of the ground state wave
function \cite{13,14}. One of the example of the MP state is the ground state
of the Affleck--Kennedy--Lieb--Tasaki (AKLT) model \cite{3}, which is
the generic model of the Haldane phase of $S=1$ chain. However, the MP wave
functions have typically short-ranged correlations and, as a rule, describe
systems with the gapped spectrum.

In \cite{pr,zp} we have studied the model with exact singlet ground state
degenerate with ferromagnetic state. This model has two different nearest
neighbor and next-nearest neighbor interactions depending on the parameter $
\nu $ and is described by the Hamiltonian
\begin{eqnarray}
H &=&-\sum_{i=1}^{M}({\bf S}_{2i-1}\cdot {\bf S}_{2i}-\frac{1}{4})-(\nu
-1)\sum_{i=1}^{M}({\bf S}_{2i}\cdot {\bf S}_{2i+1}-\frac{1}{4})  \nonumber \\
&&+\frac{\nu -1}{2\nu }\sum_{i=1}^{N}({\bf S}_{i}\cdot {\bf S}_{i+2}-\frac{1
}{4})  \label{2}
\end{eqnarray}
with periodic boundary conditions and even $N=2M$.

In fact, this Hamiltonian describe the line of transition points from the
ferromagnetic to the singlet state of the model
\begin{equation}
H=-{\sum_{i=1}^{M}}({\bf S}_{2i-1}{\bf S}_{2i}-\frac{1}{4})+J_{23}
{\sum_{i=1}^{M}}({\bf S}_{2i}{\bf S}_{2i+1}-\frac{1}{4}
)+J_{13}\sum_{i=1}^{N}({\bf S}_{i}{\bf S}_{i+2}-\frac{1}{4})
\label{1}
\end{equation}

The ground state of (\ref{1}) is ferromagnetic (singlet) at $\delta <0$ ($
\delta >0$), where $\delta =J_{13}+\frac{J_{23}}{2(1-J_{23})}$. When $\delta
=0$, the model (\ref{1}) reduces to the Hamiltonian (\ref{2}).

It has been proved in \cite{pr,zp} that singlet ground state of (\ref{2}) has
zero energy at $\nu >0$ as well as the ferromagnetic state, while all other
states have the positive energies. It has also been shown that singlet
ground state\ has double-spiral ordering (excluding some special values of
parameter $\nu $).

In \cite{pr} we have also considered the second model which is equivalent to 
special case of the spin-$\frac 12$ ladder.  This model depends on one 
parameter, has the non-degenerate singlet ground state and its ground state 
properties are similar to that of AKLT model.

The exact singlet ground state wave function of these models has a special
recurrent form:
\begin{equation}
\Psi _{0}(M)=P_{0}\Psi _{M},  \label{wf0}
\end{equation}
\begin{eqnarray}
\Psi _{M} &=&(s_{1}^{+}+\nu _{1}s_{2}^{+}+\nu _{2}s_{3}^{+}...+\nu
_{2}s_{N}^{+})(s_{3}^{+}+\nu _{1}s_{4}^{+}...+\nu _{2}s_{N}^{+})... 
\nonumber \\
&&\times (s_{2n-1}^{+}+\nu _{1}s_{2n}^{+}...+\nu
_{2}s_{N}^{+})...(s_{N-1}^{+}+\nu _{1}s_{N}^{+})\mid \downarrow \downarrow
...\downarrow \rangle   \label{wf}
\end{eqnarray}
where $s_{i}^{+}$ is the $s=\frac{1}{2}$ raising operator.
Eq.(\ref{wf}) contains $M=\frac{N}{2}$ operator multipliers and the vacuum
state $\mid \downarrow \downarrow ...\downarrow \rangle $ is the state with
all spins pointing down. The function $\Psi _{M}$ is the eigenfunction of 
$S_{z}$ with $S_{z}=0$ but it is not the eigenfunction of ${\bf S}^{2}$. 
$P_{0}$ is a projector onto the singlet state.

The models in \cite{pr} correspond to two particular cases of this wave 
function: $\nu_1=\nu_2=\nu$ for the first model (\ref{2}), and 
$\nu_1+\nu_2+1=0$ for the second model.

In this paper we consider the general case of the wave function (\ref{wf0}) 
which allows us to construct new Hamiltonian depending on two parameters 
$\nu_1$ and $\nu_2$ and having the degenerate singlet and ferromagnetic states.
The singlet ground state of this Hamiltonian has double-spiral structure 
with period of spirals equals to the system size, but 
for special values of the parameters (which include the second model in 
\cite{pr}) the singlet ground state correlations have antiferromagnetic 
character with an exponential decay and the singlet wave function in these 
cases reduces to the MP form. The interesting feature of
the model is the existence of the excited states which are asymptotically
degenerated with the ground state for special values of parameters 
in the thermodynamic limit.

The paper is organized as follows. In the Section 2 we will consider the
construction of the Hamiltonian with the exact singlet ground state and will
calculate spin correlation functions. In the Section 3 the Hamiltonian at
special values of the parameters is considered. In the Section 4 the
numerical results for the energy spectrum of the model are presented and 
Section 5 gives a brief summary.

\section{The model}

Now we will construct the Hamiltonian for which $\Psi _{0}(M)$ is the exact
ground state wave function. This Hamiltonian describes two-leg 
$s=\frac{1}{2}$ ladder with periodic boundary conditions (Fig.1) and can be 
represented in a form
\begin{figure}[t]
\unitlength1cm
\begin{picture}(8,5)
\centerline{\psfig{file=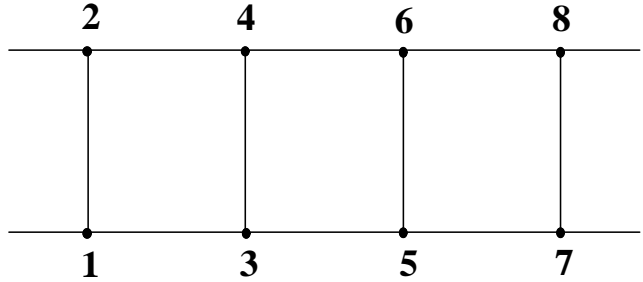,angle=-90,width=9cm}}
\end{picture}
\vspace{-20mm}
\caption{ \label{lad} The two-leg spin ladder.}
\end{figure}
\begin{equation}
H=\sum\limits_{i=1}^{M}h_{i,i+1},  \label{H}
\end{equation}
where $h_{i,i+1}$ describes the interaction between neighboring rungs. 
The spin space of two neighboring rungs consists of six multiplets: two singlet,
three triplet and one quintet. At the same time, one can check that for open
chain the wave function $\Psi _{M}$ contains only three of the six multiplets
of each pair of neighboring rungs: 
one singlet, one triplet and one quintet. The specific form of the singlet 
and triplet components present in the wave function (\ref{wf}) depends on
parameters $\nu _{1}$ and $\nu _{2}$. The Hamiltonian $h_{i,i+1}$
can be written as the sum of the projectors onto the three missing
multiplets with arbitrary positive coefficients $\lambda _{1},\,\lambda
_{2},\,\lambda _{3}$: 
\begin{equation}
h_{i,i+1}=\sum\limits_{k=1}^{3}\lambda _{k}P_{k}^{i,i+1},  \label{7}
\end{equation}
where $P_{k}^{i,i+1}$ is the projector onto the missing multiplets in the
corresponding cell Hamiltonian. Actually, the Hamiltonian $h_{i,i+1}$
contains also offdiagonal projectors between two missing triplets, but we 
use this freedom in advance to make exchange integrals on two legs of ladder 
and on each rung respectively equal, that is $J_{12}=J_{34}$ and 
$J_{13}=J_{24}$.

The wave function (\ref{wf}) is an exact wave function of the ground state
of the Hamiltonian $h_{i,i+1}$ with zero energy, because 
\begin{equation}
h_{i,i+1}|\Psi _{M}\rangle =0,\qquad i=1,...M-1  \label{8}
\end{equation}
and $\lambda _{1},\,\lambda _{2},\,\lambda _{3}$ are the excitation energies
of the corresponding multiplets.

So, $\Psi _{M}$ is the exact ground state wave function with zero energy for
the total Hamiltonian of an open ladder
\begin{equation}
H_{op}=\sum_{i=1}^{M-1}h_{i,i+1}  \label{9}
\end{equation}
\begin{equation}
H_{op}|\Psi _{M}\rangle =0  \label{10}
\end{equation}

Since the function $\Psi _M$ contains components with all possible values 
of total spin $S$ ($0\leq S\leq M$), then the ground state of open ladder 
is multiple degenerate. But it can be proved by the same way as it was made 
in \cite{pr,zp} that for a cyclic ladder (\ref{H}) only singlet and 
ferromagnetic components of $\Psi _M$ have zero energy. Therefore, for a 
cyclic ladder (\ref{H}) $\Psi _{0}(M)$ is a singlet ground state wave 
function degenerated with ferromagnetic state.
Besides, the following general statements are valid for the model (\ref{H})

1). The ground states of open ladder described by (\ref{9}) in the sector
with fixed total spin $S$ are non-degenerate and their energies are zero.

2). For cyclic ladder the ground state in the $S=0$ sector is
non-degenerate. The ground state energies for $0<S<M$ are positive.

3). The singlet ground state wave functions for open and cyclic ladders
coincide with each other.

Since the specific form of the existing and missing multiplets in the wave
function (\ref{wf}) on each two nearest neighbor spin pairs depends on the
parameters $\nu _{1}$ and $\nu _{2}$, the projectors in (\ref{7}) also
depend on $\nu _{1}$ and $\nu _{2}$. Each projector can be written in the
form 
\begin{eqnarray}
P_{k}^{1,2} &=&J_{12}^{(k)}({\bf S}_{1}\cdot {\bf S}_{2}+{\bf S}_{3}\cdot 
{\bf S}_{4})+J_{13}^{(k)}({\bf S}_{1}\cdot {\bf S}_{3}+{\bf S}_{2}\cdot {\bf 
S}_{4})+J_{14}^{(k)}{\bf S}_{1}\cdot {\bf S}_{4}+J_{23}^{(k)}{\bf S}
_{2}\cdot {\bf S}_{3}  \nonumber \\
&+&J_{1}^{(k)}({\bf S}_{1}\cdot {\bf S}_{2})({\bf S}_{3}\cdot {\bf S}
_{4})+J_{2}^{(k)}({\bf S}_{1}\cdot {\bf S}_{3})({\bf S}_{2}\cdot {\bf S}
_{4})+J_{3}^{(k)}({\bf S}_{1}\cdot {\bf S}_{4})({\bf S}_{2}\cdot {\bf S}
_{3})+C^{(k)}  \nonumber
\end{eqnarray}
and this representation is unique for a fixed value of the parameters $\nu
_{1}$ and $\nu _{2}$.

Substituting the above expressions for the projectors into Eq.~(\ref{7}), we
obtain the general form of the Hamiltonians $h_{i,i+1}$. Inasmuch as the
Hamiltonians $h_{i,i+1}$ have exactly the same form for any $i$, it suffices
here to give the expression for~$h_{1,2}$:
\begin{eqnarray}
h_{1,2}=J_{12}(A_{12}+A_{34})+J_{13}(A_{13}+A_{24})+J_{14}A_{14}+J_{23}A_{23}
\nonumber \\
+J_{1}A_{12}A_{34}+J_{2}A_{13}A_{24}+J_{3}A_{14}A_{23}
\label{e16}
\end{eqnarray}
where
\[
A_{ij}={\bf S}_{i}\cdot {\bf S}_{j}-\frac{1}{4}
\]
and all exchange integrals depend on the model parameters and the spectrum
of excited states $J_{i}=J_{i}(\nu _{1},\,\nu _{2},\,\lambda _{1},\,\lambda
_{2},\,\lambda _{3})$ as follows:
\begin{eqnarray}
J_{12} &=& -\frac{\lambda_2}{2}+\frac{\lambda_3}{2} \frac{(\nu _1 -1)^2-\nu _2
^2}{(\nu _1 -1)^2+\nu _2 ^2}   \qquad
J_{13} = -\frac{\lambda_2}{2}-\frac{\lambda_3}{2} \frac{(\nu _1 -1)^2-\nu _2
^2}{(\nu _1 -1)^2+\nu _2 ^2}  \nonumber \\
J_{14} &=& -\frac{\lambda_2}{2}-\frac{\lambda_3}{2} \frac{(\nu _2 -\nu _1
+1)^2}{(\nu _1-1)^2+\nu _2 ^2}      \qquad 
J_{23} = -\frac{\lambda_2}{2}-\frac{\lambda_3}{2} \frac{(\nu _2 +\nu _1
-1)^2}{(\nu _1-1)^2+\nu _2 ^2}     \nonumber  \\ 
J_{1}  &=& 2J_{12}- \lambda_1 \frac{(2\nu _1 - \nu _2 \nu _1 -\nu _2)(\nu _2 \nu _1 +\nu _2 -1-\nu _1
^2)}{Z}    \label{general}    \\
J_{2}  &=& 2J_{13}- \lambda_1 \frac{(\nu _1 -1)^2(\nu _2 \nu _1 +\nu _2 -1-\nu _1 ^2)}{Z}
\nonumber  \\
J_{3}  &=& J_{14}+J_{23}-\lambda_1 \frac{(\nu _1 -1)^2(2\nu _1-\nu _2 \nu _1 -\nu _2)}{Z} 
\nonumber
\end{eqnarray}
where
\[
Z = \frac{3}{4} (\nu _1 -1)^4 + \frac{1}{4} (\nu _1 +1)^2(2\nu _2 -\nu _1 -1)^2
\]
(one should keep in mind that only positive 
$\lambda _{i}$ can be substituted to these expressions). 

In general, the Hamiltonian $h_{i,i+1}$ contains all the terms presented in
(\ref{e16}), but we can simplify it 
by setting, for example, $J_{2}=J_{3}=0$ and solving equations (\ref{general}) 
for $\lambda _{1},\,\lambda _{2},\,\lambda _{3}$. All $\lambda _{i}$ turn
out to be positive in this case for any $\nu _{1}$ and $\nu _{2}$ except two
lines: $\nu _{1}=1$ and $\nu _{2}=\nu _{1}+1$, where ground state is 
multiple degenerated. The Hamiltonian $h_{i,i+1}$ in this case takes the form
\begin{equation}
h_{1,2}=J_{12}(A_{12}+A_{34})+J_{13}(A_{13}+A_{24})+J_{14}A_{14}+J_{23}A_{23}+J_{1}A_{12}A_{34}
\label{17}
\end{equation}
\begin{eqnarray*}
J_{12} &=&\frac{\nu _{1}\nu _{2}+\nu _{2}-2\nu _{1}-\nu _{2}^{2}}{2}\qquad
J_{13}=\frac{\nu _{1}\nu _{2}+\nu _{2}-\nu _{1}^{2}-1}{2} \\
J_{14} &=&\nu _{1}-\nu _{2}\qquad J_{23}=\nu _{1}(1-\nu _{2})\qquad J_{1}=4
\frac{\nu _{1}(1-\nu _{2})(\nu _{1}-\nu _{2})}{(1-\nu _{1})^{2}}
\end{eqnarray*}

The calculation of the norm of (\ref{wf0}) and the singlet ground state correlation
functions can be performed in complete analogy to the corresponding calculations
for the case $\nu _{1}=\nu _{2}$ \cite{pr,zp}.
So, a norm of $\Psi _{0}(M)$ can be written in a form
\begin{equation}
G_{M}=\langle \Psi _{0}(M)\Psi _{0}(M)\rangle =\frac{1}{2}\int_{-1}^{1}\Phi
_{M}(y)dy,  \label{g1}
\end{equation}
where $\Phi _{M}(y)$ is expanded over Legendre polynomials $P_{n}(y)$
\begin{equation}
\Phi _{M}(y)=\sum_{n=0}^{M}c_{n}(M)P_{n}(y)  \label{sum}
\end{equation}

The coefficients $c_{n}(l)$ are defined by the recurrent equation
\begin{eqnarray}
c_{n}(l+1) &=&\frac{n}{2n-1}\frac{\left[ \nu _{2}(n-1)+\nu _{1}+1\right] ^{2}
}{2}c_{n-1}(l)+\frac{\nu _{2}^{2}(n^{2}+n)+(\nu _{1}-1)^{2}}{2}c_{n}(l) 
\nonumber \\
&&+\frac{n+1}{2n+3}\frac{\left[ \nu _{2}(n+2)-\nu _{1}-1\right] ^{2}}{2}
c_{n+1}(l)  \label{g8}
\end{eqnarray}
with initial condition $c_{0}(0)=1$ and $c_{n}(l)=0$ at $n>l$.

The appropriate calculations result in the expression for spin correlation
functions at $N\rightarrow \infty $
\begin{equation}
\left\langle {\bf S}_{1}{\bf S}_{2l+1}\right\rangle =
\left\langle {\bf S}_{2}{\bf S}_{2l+2}\right\rangle =
\frac{1}{4}\cos \left( \frac{4\pi l}{N}\right)   \label{36} 
\end{equation}
\begin{equation}
\left\langle {\bf S}_{1}{\bf S}_{2l+2}\right\rangle =
\frac{1}{4}\cos \left( \frac{4\pi l}{N}+\triangle \varphi \right) \label{37}
\end{equation}

These equations mean that the spiral on each leg with pitch angle $\frac{
4\pi }{N}$ is formed and the shift angle between spirals on the upper and
the lower legs is $\triangle \varphi =\frac{4\pi }{N}\frac{\nu _{1}-1}
{\nu _{2}}$. So, there is just one full rotation of the spin over the 
length of the ladder, independent of the size of the system and for fixed 
$l<<N$ at $N\rightarrow \infty $ two spins on the ladder are parallel.

\section{Special cases}

There are special values of the parameters $\nu _{1}$ and $\nu _{2}$ for
which Eqs.(\ref{36}) and (\ref{37}) are not valid. These values can be
determined from Eq.(\ref{g8}), when the coefficients of $c_{n-1}(l)$ or $
c_{n+1}(l)$ equal to zero. These conditions lead to following equations
\begin{equation}
\nu _{2}(n-1)+\nu _{1}+1=0  \label{l1}
\end{equation}
\begin{equation}
-\nu _{2}(n+1)+\nu _{1}+1=0  \label{12}
\end{equation}

The special lines (\ref{l1}) and (\ref{12}) on the ($\nu _{1},\ \nu _{2}$)
plane are shown in Fig.2. At $\nu _{1}=\nu _{2}$ Eqs.(\ref{l1}) and (\ref{12})
give special points of the model (\ref{2}) \cite{pr,zp}.
\begin{figure}[t]
\unitlength1cm
\begin{picture}(12,7)
\centerline{\psfig{file=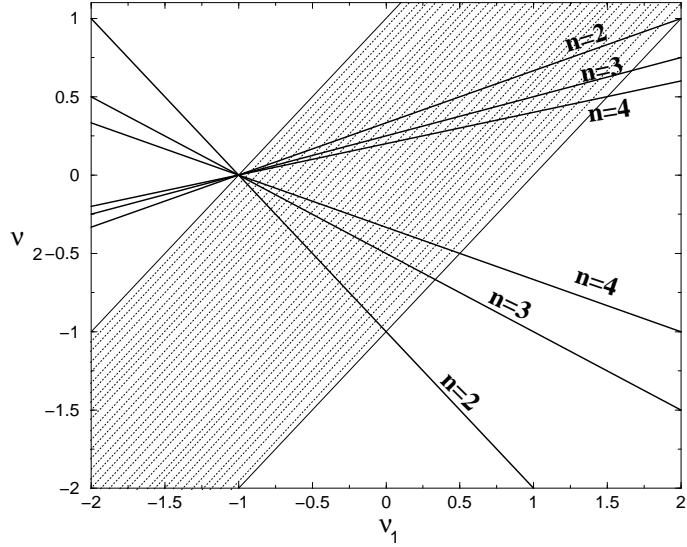,angle=-90,width=9cm}}
\end{picture}
\caption{ \label{nunu} 
($\nu _1$, $\nu _2$) plane. Thick lines are special cases of the model.
Each value of $n$ corresponds to a pair of equivalent lines.
Due to the symmetry (21) it is sufficient to consider only the
shaded area of ($\nu _1$, $\nu _2$) plane.}
\end{figure}

We note that there is a symmetry transformation
\begin{equation}
\nu _{1}\rightarrow \frac{\nu _{2}-1}{\nu _{2}-\nu _{1}}\qquad \nu
_{2}\rightarrow \frac{\nu _{2}}{\nu _{2}-\nu _{1}}  \label{tr}
\end{equation}
which maps lines (\ref{l1}) and (\ref{12}) onto each other and, therefore,
there is pair of equivalent lines for each value of $n$. The transformation
(\ref{tr}) does not change the wave function (\ref{wf0}) as well as the Hamiltonian
(\ref{17}) apart from a change in energy scale by a factor $\left( \nu
_{2}-\nu _{1}\right) ^{2}$. So, it is sufficient to consider the region on ($
\nu _{1}$,$\nu _{2})$ plane restricted by the inequality $\left| \nu
_{2}-\nu _{1}\right| \leq 1$ (see Fig.2).

For $\nu _{1}$ and $\nu _{2}$ defined by Eqs.(\ref{l1}) or (\ref{12}) $\Phi
_{M}(y)$ contains only $n$ terms and the wave function $\Psi _{M}$ contains
only $n$ multiplets rather than $(M+1)$ as it does in generic case. It can be
shown \cite{pr}\ that wave function $\Psi _{0}(M)$ for these special cases\
can be written in MP form:
\begin{equation}
\Psi _{0}(M)=Tr\,\left( D_{1,2}\ D_{3,4}\,\ldots \,D_{N-1,N}\right) ,
\label{s2}
\end{equation}
where $D=T+uS$ is the $n\times n$ matrix describing states of corresponding
spin pair. Singlet state matrix is
\begin{equation}
S=I\ \left| s\right\rangle   \label{s4}
\end{equation}
where $I$ is identity matrix and $\left| s\right\rangle $ is the singlet
state. Triplet state matrix $T$\ is expressed by Clebsch-Gordan coefficients 
$C_{m_{1},m_{2}}=\left\langle \left( 1,m_{1}\right) \left( j,m_{2}\right) \
|\ \left( j,m_{1}+m_{2}\right) \right\rangle $ as follows:
\begin{equation}
T=\frac{1}{C_{0,j}}\left( 
\begin{array}{ccccc}
C_{0,j}\left| 0\right\rangle  & C_{1,j-1}\left| 1\right\rangle  & 0 & 0 & 0
\\ 
C_{-1,j}\left| -1\right\rangle  & C_{0,j-1}\left| 0\right\rangle  & . & 0 & 0
\\ 
0 & . & . & . & 0 \\ 
0 & 0 & . & . & C_{1,-j}\left| 1\right\rangle  \\ 
0 & 0 & 0 & C_{-1,-j+1}\left| -1\right\rangle  & C_{0,-j}\left|
0\right\rangle 
\end{array}
\right) ,  \label{s5}
\end{equation}
where $j=\frac{n-1}{2}$ and $\left| \sigma \right\rangle $ is the triplet
state with $S^{z}=\sigma $. The parameter $u$ is defined by
expression
\begin{equation}
u=\frac{\nu _{1}-1}{\nu _{2}\left( n-1\right) }  \label{s6}
\end{equation}

Exact calculation of the correlators in these cases using standard transfer
matrix technique results in
\begin{eqnarray}
\left\langle {\bf S}_{1}{\bf S}_{2}\right\rangle  &=&\frac{1}{4}-\frac{u^{2}
}{\omega _{1}}  \nonumber \\
\left\langle {\bf S}_{i}{\bf S}_{i+2l}\right\rangle  &=&(1+2z)\frac{
u^{2}-z^{2}}{\omega _{1}^{2}}\left( \frac{\omega _{2}}{\omega _{1}}\right)
^{l-1}  \nonumber \\
\left\langle {\bf S}_{2i+1}{\bf S}_{2i+2l+2}\right\rangle  &=&-(1+2z)\frac{
(u+z)^{2}}{\omega _{1}^{2}}\left( \frac{\omega _{2}}{\omega _{1}}\right)
^{l-1}  \nonumber \\
\left\langle {\bf S}_{2i+2}{\bf S}_{2i+2l+1}\right\rangle  &=&-(1+2z)\frac{
(u-z)^{2}}{\omega _{1}^{2}}\left( \frac{\omega _{2}}{\omega _{1}}\right)
^{l-1}  \label{corr}
\end{eqnarray}
where we use notations
\[
z=\frac{1}{n-1},\qquad \omega _{1}=1+2z+u^{2},\qquad \omega _{2}=\omega
_{1}-4z^{2}
\]

In the particular case of zero singlet weight, $u=0$, when spins on each
rung form a local triplet, correlation functions (\ref{corr}) coincide with
those obtained in \cite{13,18}.

According to Eqs.(\ref{corr}) the singlet ground state has collinear or
stripe spin structure, i.e. spin-spin correlations are ferromagnetic along
legs and antiferromagnetic between them (Fig.3).
\begin{figure}[t]
\unitlength1cm
\begin{picture}(8,5)
\centerline{\psfig{file=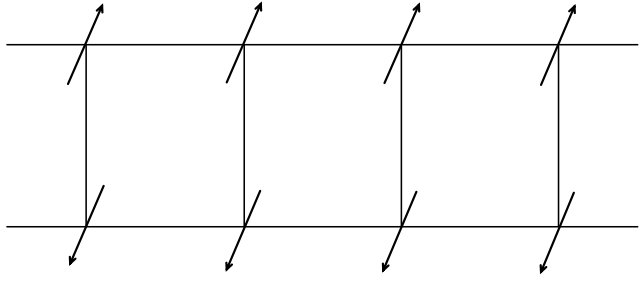,angle=-90,width=9cm}}
\end{picture}
\vspace{-20mm}
\caption{ \label{str} Stripe spin structure on the ladder model.
}
\end{figure}

These correlations have an exponential decay for finite value $n$ and the
correlation length $r_{c}$ is
\begin{equation}
r_{c}=2\ln ^{-1}\left| \frac{\omega _{1}}{\omega _{2}}\right|   \label{rc}
\end{equation}

For $n\rightarrow \infty $ there is a magnetic order $m$:
\[
\left\langle {\bf S}_{i}{\bf S}_{i+2l}\right\rangle =
- \left\langle {\bf S}_{i}{\bf S}_{i+1+2l}\right\rangle =
m^{2} \qquad l\rightarrow \infty ,\qquad m=\frac{u}{1+u^{2}}
\]

When $u=\pm 1$ the magnetic order is equal to the classical value $1/2$.

It is interesting to note that $r_{c}$ diverges for finite $n$ when $
u\rightarrow \infty $ ($(\nu _{1},\nu _{2})\rightarrow (-1,0)$) but the
prefactors in (\ref{corr}) vanish in this case. As it is shown on Fig.2 this
is the point where all special lines are intersected and the wave function $
\Psi _{0}(M)$ is a simple product of the singlet pairs
\begin{equation}
\Psi _{0}(M)=(s_{1}^{+}-s_{2}^{+})\ ...\ (s_{N-1}^{+}-s_{N}^{+})\mid
\downarrow \downarrow ...\downarrow \rangle   \label{singlet}
\end{equation}

We note that the wave function (\ref{wf0}) show double-spiral ordering for
all values of $\nu _{1}$ and $\nu _{2}$ excluding the special lines. The
crossover between spiral and strip states occurs in the exponentially small
(at $N\rightarrow \infty $) vicinity of the special lines.

Now we should make the following remark. There is one particular case $\nu
_{1}+\nu _{2}+1=0$ ($n=2$), which was considered in \cite{pr}, when on each
two neighboring rungs the wave function $\Psi _{M}$ contains only one singlet
and one triplet and does not contain quintet. Therefore, in this 
case the cell Hamiltonian can be written in the form \cite{JETP}
\begin{equation}
H_{i,i+1}=\sum\limits_{k=1}^{4}\lambda _{k}P_{k}^{i,i+1},  \label{q13}
\end{equation}
where $P_{4}^{i,i+1}$ - is a projector onto the quintet state. If $\lambda
_{1},\,\lambda _{2},\,\lambda _{3},\,\lambda _{4}>0$, wave function $\Psi
_{0}(M)$ is non-degenerate singlet ground state wave function for the
Hamiltonian (\ref{q13}). In this case all four-spin interactions can be
excluded by setting $J_{1}=J_{2}=J_{3}=0$ and then we arrive at the
Hamiltonian
\begin{eqnarray}
H_{i,i+1} &=&J_{12}\left[ ({\bf S}_{2i-1}\cdot {\bf S}_{2i}-\frac{1}{4})+(
{\bf S}_{2i+1}\cdot {\bf S}_{2i+2}-\frac{1}{4})\right]   \nonumber \\
&&+2J_{13}\left[ ({\bf S}_{2i-1}\cdot {\bf S}_{2i+1}-\frac{1}{4})+({\bf S}
_{2i}\cdot {\bf S}_{2i+2}-\frac{1}{4})\right]   \label{h22} \\
&&+2J_{14}({\bf S}_{2i-1}\cdot {\bf S}_{2i+2}-\frac{1}{4})+2J_{23}({\bf S}
_{2i}\cdot {\bf S}_{2i+1}-\frac{1}{4})  \nonumber
\end{eqnarray}
and all exchange integrals $J_{ij}$ depend on one model parameter. The
explicit form of $J_{ij}$ was found in \cite{pr}. It was shown that for this
model $\Psi _{0}(M)$ is non-degenerate singlet ground state wave function
with exponentially decaying spin correlations and there is an energy gap.

In other special cases we can also construct Hamiltonians for which $\Psi
_{0}(M)$ is non-degenerate singlet ground state wave function. But we have
to introduce more distant interactions. For example, such a model for $n=3$
would contain interactions between next-nearest neighbor spin pairs
\begin{equation}
H=\sum\limits_{i=1}^{M}H_{i,i+1,i+2}  \label{Hs1}
\end{equation}

For the point ($\nu _{1}=1,\ \nu _{2}=-1$), when spins on each rung form a
local triplet, one of the possible Hamiltonians can be written as
\begin{eqnarray}
H_{1,2,3}
&=&A_{12}+A_{23}+38A_{13}+6A_{13}^{2}-(A_{12}^{2}+A_{23}^{2})(A_{13}+\frac{7
}{4})  \nonumber \\
&&-(A_{13}+\frac{7}{4})(A_{12}^{2}+A_{23}^{2})+52  \label{52}
\end{eqnarray}
where
\[
A_{ij}={\bf L}_{i}\cdot {\bf L}_{j}-1
\]
and ${\bf L}_{i}={\bf s}_{2i-1}+{\bf s}_{2i}$ is the $S=1$ operator.

The ground state wave function for the Hamiltonian (\ref{52}) has the
matrix-product form (\ref{s2}) with
\[
D=\left( 
\begin{array}{ccc}
\left| 0\right\rangle  & -\left| 1\right\rangle  & 0 \\ 
\left| -1\right\rangle  & 0 & -\left| 1\right\rangle  \\ 
0 & \left| -1\right\rangle  & -\left| 0\right\rangle 
\end{array}
\right) 
\]

\section{Spectrum of the model}

Generally, the excitation spectrum of the model (\ref{H}, \ref{17}) can not
be calculated exactly. It is clear that this spectrum is gapless because, for
example, the one-magnon energy is $\sim N^{-4}$ at $N\rightarrow \infty $.
Moreover, the lowest singlet excitation is gapless as well. For the model 
(\ref{H}, \ref{17}) lying on the special lines this fact can be established 
from the following consideration. For the simplicity we consider the
case $\nu _{1}=\nu _{2}=\nu =\frac{1}{n}$. We choose the variational wave
function of the excited singlet state at $\nu _{1}=\nu _{2}=\nu $ as
\[
\Psi _{s}(\nu ,\delta ,M)=\frac{1}{\sqrt{1-c^{2}(\delta ,M)}}\left[ c(\delta
,M)\ \Psi _{0}(\nu ,M)-\Psi _{0}(\nu +\delta ,M)\right] ,
\]
where $\Psi _{0}(\nu ,M)$ and $\Psi _{0}(\nu +\delta ,M)$ are the normalized
singlet ground state wave functions of (\ref{H}) with zero energies at $\nu
_{1}=\nu _{2}=\nu $ and $\nu _{1}=\nu _{2}=\nu +\delta $ (the point $\nu
_{1}=\nu _{2}=\nu +\delta $ on ($\nu _{1},\nu _{2}$) plane does not belong
to the special line). The functions $\Psi _{s}(\nu ,\delta ,M)$ and $\Psi
_{0}(\nu ,M)$ are orthogonal and $c(\delta ,M)$ is the overlap of $\Psi
_{0}(\nu ,M)$ and $\Psi _{0}(\nu +\delta ,M)$
\[
c(\delta ,M)=\left\langle \Psi _{0}(\nu ,M)\ |\ \Psi _{0}(\nu +\delta
,M)\right\rangle 
\]

It can be shown that
\begin{eqnarray}
c^2 (\delta ,M) \sim \frac{1}{1+\delta ^2 (M!)^2 e^{O(M)}}
\quad at\quad M\rightarrow \infty \label{d(N)}
\end{eqnarray}

Eq.(\ref{d(N)}) follows from the fact that $\left\langle \Psi _{0}(\nu ,M)\
|\ \Psi _{0}(\nu ,M)\right\rangle $ and $c(\delta ,M)$ is represented by the
sum of $n$ terms in (\ref{sum}) while $\left\langle \Psi _{0}(\nu +\delta
,M)\ |\ \Psi _{0}(\nu +\delta ,M)\right\rangle $ contains $M$ terms.

The variational energy calculated with respect to $\Psi _{s}(\nu ,\delta ,M)$
is
\begin{eqnarray*}
E_{s} &=&\left\langle \Psi _{s}(\nu ,\delta ,M)\ |\ H(\nu ,M)\ |\ \Psi
_{s}(\nu ,\delta ,M)\right\rangle  \\
&=&\frac{1}{1-c^{2}(\delta ,M)}
\left\langle \Psi _{0}(\nu +\delta ,M)\ |\
H(\nu ,M)\ |\ \Psi _{0}(\nu +\delta ,M)\right\rangle  \\
&\sim &\frac{1+\delta ^2 (M!)^2 }{\delta ^2 (M!)^2 }
\left\langle \Psi _{0}(\nu +\delta ,M)\ |\
(-\delta \frac{dH(\nu ,M)}{d\nu })\ |\ \Psi _{0}(\nu +\delta ,M)\right\rangle \\
&\sim &\frac{1+\delta ^2 (M!)^2 }{\delta (M!)^2 }
\end{eqnarray*}

And minimization of $E_{s}$ over the function $\delta (M)$ leads to 
\begin{eqnarray*}
\delta \sim e^{-M\ln M +O(M)} \qquad
E_{s} \sim e^{-M\ln M +O(M)}
\end{eqnarray*}

This consideration can be easily extended to the points
$\nu _{1}\neq \nu _{2}$ on the special lines. But it is not valid for the
parameters $\nu _{1}$ and $\nu _{2}$ which are out of special lines.
We performed the numerical diagonalization of finite ladders for various
parameters $\nu _{1}$ and $\nu _{2}$. The energies of lowest singlet and
triplet states of (\ref{H}) are shown in Fig.4 as a function of $N$ for
parameters corresponding to different types of the ground state.
Figure 4 shows that the exponential degeneracy possibly takes place for
all parameters $\nu _{1}$ and $\nu _{2}$, but we can not confirm it strictly.
\begin{figure}[t]
\unitlength1cm
\begin{picture}(12,7)
\centerline{\psfig{file=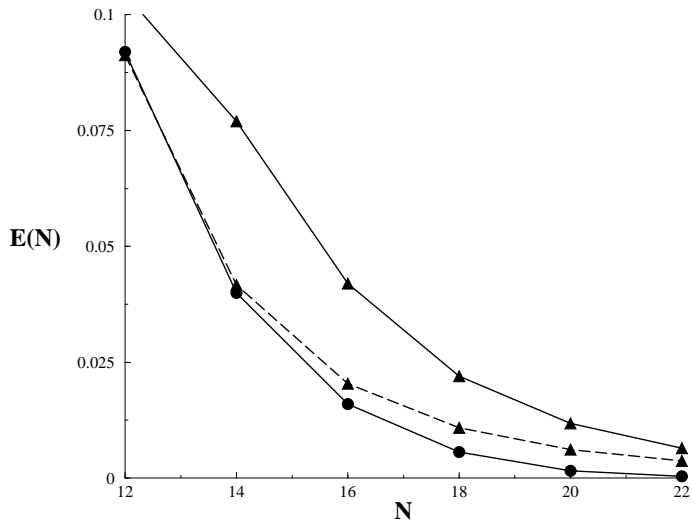,angle=-90,width=9cm}}
\end{picture}
\caption{ \label{en} 
Dependences of energy of the lowest singlet (circles) and triplet
(triangles) excited states on the system size $N$ at the points
$\nu _1 = \nu _2 = 1/2$ (solid lines) and $\nu _1 = \nu _2 = 3/4$ 
(dashed line).
The singlet excited state energy for $\nu _1=\nu _2=3/4$ is out of scale of
this figure but has similar dependence on $N$.
}
\end{figure}

Thus, on the special lines the ground state of considered model is
asymptotically degenerated at the thermodynamic limit. It is not clear
if the degeneracy is exponentially large or not.

So far, we have considered the models with degenerate singlet and
ferromagnetic states. Now we discuss the phase diagram of the zigzag chain
model given by (\ref{1}). The line of transition points from the
ferromagnetic to singlet state is described by the one parametric
Hamiltonian (\ref{2}) (or by (\ref{17}) at $\nu _{1}=\nu _{2}$).

The exact ground state in the singlet phase (Fig.5) is generally unknown. But
it is interesting to note that the ground state on the line $J_{13}=-1/2$ is
the product of singlets on ladder diagonals (2,3), (4,5), ... as in the
point $J_{13}=-J_{23}=-1/2$ on the transition line. The spectrum of (\ref{1})
on the transition line is gapless. There are some regions on the plane ($
J_{13}$, $J_{23}$) which were studied by different approximation methods.
\begin{figure}[tbp]
\unitlength1cm
\begin{picture}(13,8)
\centerline{\psfig{file=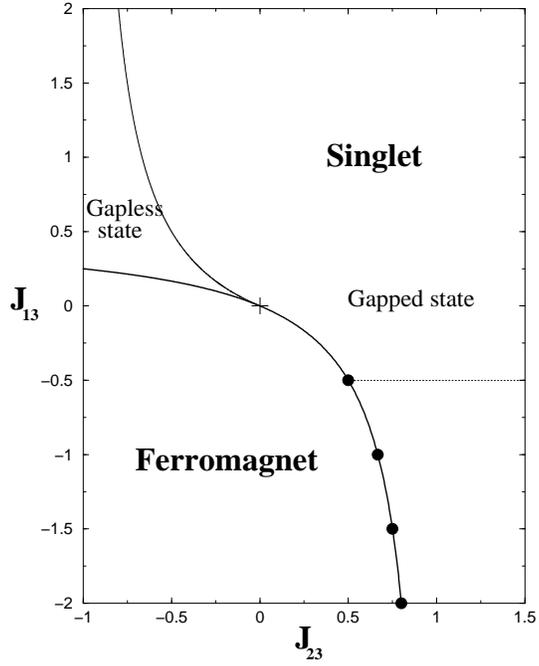,angle=0,width=7cm}}
\end{picture}
\vspace{10mm}
\caption{ \label{jj} 
Phase diagram of the zigzag chain model (2). The thick solid line is the 
boundary between the ferromagnetic and singlet phases. Circles
correspond to the special cases of the model. The thin solid line
denotes the heuristic boundary between gapped and gapless phases.
On the dotted line the ground state is a product of singlet pairs
$[2,3][4,5]...$.
}
\end{figure}

At $J_{13}=0$ and $0<J_{23}<1$ the model (\ref{1}) reduces to the
alternating Heisenberg chain studied in \cite{19}. The lowest excitation
is the triplet and there is the gap. At $J_{23}=0$ and $J_{13}>0$ the model
(\ref{1}) reduces to the spin ladder with antiferromagnetic interactions
along legs and the ferromagnetic interactions on rungs. It is evident that
there is a gap at $J_{13}\ll 1$ (in this case the model is equivalent to the
spin $S=1$ Heisenberg chain). It was shown in \cite{20} that the gap
exists at $J_{13}\gg 1$. At $J_{23}=-1$ and $J_{13}\gg 1$ the spectrum is
gapless according to the results of \cite{21}. 

We have calculated the
first singlet and triplet excitation at $J_{13}=-1/2$ and $1/2<J_{23}<1$ by
the numerical diagonalization of the finite ladders. As it can be seen from
Fig.6 the gap is closed on the transition line. So, we expect that the phase
diagram of the model (\ref{1}) has the form shown in Fig.5.
\begin{figure}[tbp]
\unitlength1cm
\begin{picture}(12,7)
\centerline{\psfig{file=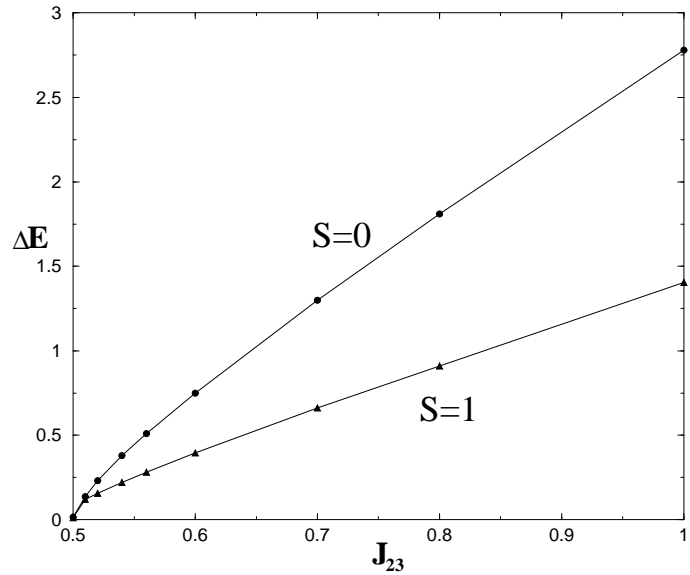,angle=-90,width=9cm}}
\end{picture}
\vspace{10mm}
\caption{ \label{ej}
Dependence of singlet-singlet and singlet-triplet energy gap on $J _{23}$
along the line $J _{13}=-1/2$ (dotted line in Fig.5). Calculation was
made for finite chain with $N=20$.
}
\end{figure}

\section{Summary}

We have constructed the spin ladder model with ferro- and antiferromagnetic
interactions between spins on neighboring rungs. The model has exact singlet
ground state degenerated with ferromagnetic state. The spin correlators in
the singlet ground state show double-spiral ordering with period of spirals 
equals to the system size. However, for special
values of the parameters spin correlators in the singlet state have
exponential decay and in these cases the singlet ground state wave function
can be represented in the MP form. The spectrum of the model is gapless and
there is asymptotic degeneracy of the ground state for special values of 
the parameters at the thermodynamic limit.

The singlet ground state wave function has the recurrent form (\ref{wf0})
and depends on two parameters. This function can be further generalized. For
example, we can take $\Psi _{0}(M)$ in a form
\[
\Psi _{0}(M)=P_{0}\Psi _{M}
\]
where $\Psi _{M}$ is the product of alternating multipliers
\[
\Psi _{M}=(s_{1}^{+}+\nu _{1}s_{2}^{+}+\nu _{2}s_{3}^{+}...)(s_{3}^{+}+\nu
_{3}s_{4}^{+}+\nu _{4}s_{5}^{+}...)(s_{5}^{+}+\nu _{1}s_{6}^{+}+\nu
_{2}s_{7}^{+}...)...\mid \downarrow \downarrow ...\downarrow \rangle 
\]
with the condition
\[
\frac{1+\nu _{1}}{\nu _{2}}=\frac{1+\nu _{3}}{\nu _{4}}
\]

The Hamiltonian of the model for which $\Psi _{0}(M)$ is the singlet ground
state has two rungs in the elementary cell. This Hamiltonian can be taken in
the form containing the interactions between neighboring rungs only and
without four spin terms.

\section{Acknowledgements}

D.D. is grateful to Max-Planck-Institut fur Physik Komplexer Systeme,
where this work was completed, for kind hospitality. This work was
supported by RFFR (Grants N97-03-33727 and N96-15-97492). A.A.O.
appreciates the support over the Grant N872 of the ISCT.


\begin{thebibliography}{99}

\bibitem{1}E.~Dagotto, Int. J. Mod. Phys. B {\bf 5}, (1991) 907.

\bibitem{2} C.~K. Majumdar and D.~K. Ghosh, J. Math. Phys. (N.Y.) {\bf 10},
 (1969) 1388, 1399.

\bibitem{3} I.~Affleck, T.~Kennedy, E.~H. Lieb, and H.~Tasaki, Phys. Rev.
Lett. {\bf 59}, (1987) 799; Commun. Math. Phys. {\bf 115}, (1988) 477.

\bibitem{10}  I.Bose, S.Gayen, Phys. Rev. B{\bf 48} (1993) 10653.

\bibitem{4} H.Niggemann, J.Zittarz, Z. Phys. B {\bf 101} (1996) 289.

\bibitem{8}  A.K.Kolezhuk, H.-J.Mikeska, S.Yamamoto, Phys. Rev. B{\bf 
55} (1997) R3336.

\bibitem{9}  A.K.Kolezhuk, H.-J.Mikeska, Phys. Rev. B{\bf 
56} (1997) R11380.

\bibitem{11} Z.Weihong, V.Kotov, J.Oitmaa, e-print condmat/9711006.

\bibitem{9a}  A.K.Kolezhuk, H.-J.Mikeska, Europ. Phys. Journ. B{\bf 
5} (1998) 543;  Phys. Rev. Lett.{\bf 80} (1998) 2709.

\bibitem{12} T.~Kennedy, E.~H. Lieb, and H.~Tasaki, J. Statist. Phys. {\bf%
53}, (1988) 383.

\bibitem{13} M.Fannes, B.Nachtergaele, R.F.Werner, Commun. Math. Phys.
{\bf 144} (1992) 443.

\bibitem{14} A.~Klumper, A.~Schadschneider, J.~Zittartz, Z. Phys. B.
{\bf 87}, 281 (1992); Europhys. Lett. {\bf 24}(4), 293 (1993); 

\bibitem{pr}  D.V.Dmitriev, V.Ya.Krivnov, A.A.Ovchinnikov, Phys. Rev. B{\bf 
56} (1997) 5985.

\bibitem{zp}  D.V.Dmitriev, V.Ya.Krivnov, A.A.Ovchinnikov, Z.Phys. B{\bf 103}
(1997) 193.

\bibitem{JETP}  D.V.Dmitriev, V.Ya.Krivnov, A.A.Ovchinnikov, JETP {\bf 88},
(1999) 138.

\bibitem{18} W.D.Freitag, E.Muller-Hartmann, Z.Phys. B {\bf 83} (1991) 381.

\bibitem{19} K.Hida, Phys. Rev. B {\bf 45}, (1992) 2207.

\bibitem{20} D.G.Shelton, A.A.Nersesyan, A.M.Tsvelik, Phys. Rev. B {\bf 53},
(1996) 8521.

\bibitem{21} S.~R. White and I.~Affleck, Phys. Rev. B {\bf 54}, (1996) 9862.

\end{thebibliography}
\end{document}